\documentclass[aps,prc,twocolumn,showpacs,floatfix,nofootinbib]{revtex4-1}
\usepackage{amsmath,graphicx,color}
\usepackage[colorlinks,linkcolor=blue,anchorcolor=blue,urlcolor=blue,citecolor=blue]{hyperref}
\usepackage{subfigure}

\def\E{{\mathcal E}}

\def\Eq#1{Eq.~(\ref{#1})}

\def\App#1{Appendix~\ref{#1}}
\def\Fig#1{Fig.~\ref{#1}}
\def\Sect#1{Section~\ref{#1}}
\def\Ref#1{Ref.~\cite{#1}}
\def\Tab#1{Table~\ref{#1}}

\def\bra{\langle}
\def\ket{\rangle}

\def\ie{\emph{i.e.}}

\def\be{\begin{equation}}
\def\ee{\end{equation}}
\def\bea{\begin{eqnarray}}
\def\eea{\end{eqnarray}}

\begin{document}

\title{Hydrodynamic response in simulations within a multiphase transport model}
\author{De-Xian Wei}
\author{Xu-Guang Huang}
\email{huangxuguang@fudan.edu.cn}
\affiliation{Department of Physics and Center for Field Theory and Particle Physics, Fudan University, Shanghai, 200433, China}
\affiliation{Key Laboratory of Nuclear Physics and Ion-beam Application (MOE), Fudan University, Shanghai 200433, China}
\author{Li Yan}
\email{li.yan@physics.mcgill.ca}
\affiliation{Department of Physics, McGill University 3600 rue University Montr\'eal, QC Canada H3A 2T8}

\begin{abstract}

We carry out simulations using a multiphase transport (AMPT) model to describe the observed flow signatures
in $\sqrt{s_{NN}}=2.76$ TeV Pb-Pb collisions. Especially, we calculate
the flow fluctuations of $v_2$ in terms of cumulant ratios and the
standardized skewness. Based on event-by-event AMPT simulations,
we study the linear and cubic response relation between $v_2$ and $\varepsilon_2$.
We found that the observed response relation is compatible to what has
been noticed in hydrodynamic modelings, with similar dependence on shear viscosity.
Besides, this response relation is not sensitive to nonflow effects.

\end{abstract}
\maketitle

\section{INTRODUCTION}

One remarkable achievement in high-energy heavy-ion experiments is the
creation of a fluid-like quark-gluon system: the quark-gluon plasma (QGP).
At energies available at the BNL Relativistic Heavy Ion Collider (RHIC) and the CERN
Large Hadron Collider (LHC), it has been realized that
the physics of the created QGP medium can be understood in
terms of relativistic viscous hydrodynamics (see \cite{Jeon:2015dfa} for a
recent review), with extremely small dissipative corrections.
For instance, simulations based on 
hydrodynamic modelings of the QGP medium evolution provide so far
the best description of the so-called harmonic flow
$V_n$, and correlations and fluctuations of
these flow~\cite{Gale:2012rq,Chattopadhyay:2017bjs,McDonald:2016vlt,Niemi:2015voa}, given
an input of the specific shear viscosity (the ratio of shear viscosity to entropy
density) close to a lower theoretical bound $\eta/s=\hbar/4\pi k_B$~\cite{Kovtun:2004de}.

Harmonic flow $V_n$ characterizes the azimuthal anisotropy of the generated particle
spectrum in momentum space~\cite{Ollitrault:1992bk,Alver:2010gr}.
For each collision event, with respect to the probability distribution of the
generated particles in azimuthal angle, $f(\phi_p)$, one defines $V_n$
as
\be
\label{eq:vn}
V_n = v_n e^{in\Psi_n} \equiv \int \frac{d\phi}{2\pi} e^{in\phi_p} f(\phi_p)\,. 
\ee
The parameter $n$ denotes harmonic order, with $n=2$ corresponding to elliptic flow,
$n=3$ corresponding to triangular flow, etc.
Note that $V_n$ is a complex quantity by definition, which
depends in principle on particle species, transverse momentum,
pseudorapidity, etc. On an event-by-event basis,
both its magnitude $v_n$, and phase $\Psi_n$ fluctuate.
Correlations and fluctuations of $V_n$ determine all kinds of the measured
flow signatures in experiments, such as the cumulants of flow and event-plane
correlations~\cite{Aad:2014fla}.

Hydrodynamic modelings of heavy-ion collisions have led to a set of
response relations between $V_n$ and the fluctuating
initial-state geometry of the colliding systems. More precisely, these
relations are written in terms of initial state eccentricity $\E_n$,
which is defined with respect to the initial-state energy density profile $\rho(\vec x_\perp, \tau_0)$
as~\cite{Teaney:2010vd}
\footnote{
For $n=1$, the dipolar anisotropy is defined as
\[
\E_1=\varepsilon_1 e^{in\Phi_1}\equiv -{\int d^2\vec x_\perp \rho(\vec x_\perp, \tau_0) r^3 e^{i\phi}
\over
\int d^2\vec x_\perp r^3\rho(\vec x_\perp, \tau_0)}
\]
}
\be
\label{eq:en}
\E_n=\varepsilon_n e^{in\Phi_n}\equiv -{\int d^2\vec x_\perp \rho(\vec x_\perp, \tau_0) r^n e^{in\phi}
\over
\int d^2\vec x_\perp r^n\rho(\vec x_\perp, \tau_0)}\,.
\ee
Note that, since $|\E_n|=\varepsilon_n<1$ by definition, harmonic flow $V_n$ can be expanded
with respect to $\E_n$, and a series of response relation can be obtained.
A recent review on these
response relations can be found in \Ref{Yan:2017ivm}. Although the response relation for $V_2$ and similar
response relations for higher order flow are empirical, based
on event-by-event hydrodynamic simulations, they are conceptually compatible
with the physics of hydrodynamic response theory. In particular, one may understand these response as
evolution of long-wavelength hydrodynamic modes, in the way that the response coefficients
solely depend on the medium dynamical properties.
Therefore, in one selected centrality class where system multiplicity is
roughly constant, fluctuations of these response coefficients
can be ignored.
In recent experiments, information
on the flow correlations and fluctuations have been acquired with
high precision, from which, 
hydrodynamic response relations can be examined~\cite{Acharya:2017zfg,Yan:2015jma,Giacalone:2016eyu,Giacalone:2016afq}.

Although hydrodynamic response relations are well established in hydrodynamic modelings,
it is of interest to analyze these relations beyond hydrodynamics. In particular,
one notices that
nonflow effects result in additional event-by-event
fluctuations, which are not characterized in hydrodynamics.
In this work, by simulations based on a multiphase transport (AMPT)
model~\cite{Lin:2004en},
we reexamine the hydrodynamic
response relation between $V_2$ and $\E_2$. This paper is organized as
follows: In \Sect{sec:sec2} we briefly describe the AMPT model and parameters used in
the present simulations. Flow signatures are obtained correspondingly by
correlating generated particles. In particular, fluctuations of $v_2$ are
studied in terms of flow cumulants. The hydrodynamic response relation for $v_2$
is detailed in \Sect{sec:sec3}, where we emphasize its dependence
on shear viscosity and nonflow effects. Throughout this paper, our results and
analyses are mostly obtained based on
AMPT simulations with respect to Pb-Pb collisions at $\sqrt{s_{NN}}=2.76$ TeV at the LHC.
Similar results of the recent $\sqrt{s_{NN}}=5.02$ TeV Pb-Pb
collisions are presented in \App{app1}. We will use natural unit $k_B=c=\hbar=1$.

\section{AMPT AND HEAVY-ION COLLISIONS}
\label{sec:sec2}

The AMPT model is a hybrid model in which QGP evolution in heavy-ion collisions
is described by parton scatterings~\cite{Lin:2004en}. In the AMPT model, the initial-state particle
distributions are generated by the HIJING model~\cite{PhysRevD.44.3501}. For the current study, string
melting is considered so that the produced hadrons from HIJING model are further converted
into valence quarks and antiquarks. Right before parton scatterings, we record the generated
energy density profile of the system $\rho(\vec x,\tau_0)$, as the initial state of medium
evolution. Initial state eccentricities of each event are then calculated with respect
to \Eq{eq:en}. Parton scatterings, and accordingly the space-time evolution of
QGP, are determined via ZPC parton cascade model~\cite{ZHANG1998193}, with the differential cross section
\be
\frac{d\sigma}{dt}\approx\frac{9\pi\alpha_s^2}{2(t-\mu^2)^2}\,.
\ee
In the above equation, $\alpha_s$ is the strong coupling constant, $t$ is the Mandelstam variable,
and $\mu$ is the screening mass in the partonic system. These parameters are adjustable according
to colliding systems so that measurable quantities in experiments, such as total yields,
elliptic flow $v_2$,
and two-pion correlations, can be reproduced. For later convenience, we also notice
the following relation~\cite{Xu:2011fi},
\be
\label{eq:etas}
\frac{\eta}{s}\approx \frac{3\pi}{40\alpha_s^2}
\frac{1}{\left(9+\frac{\mu^2}{T^2}\right)\ln\left(\frac{18+\mu^2/T^2}{\mu^2/T^2}\right)-18},
\ee
which allows one to estimate the specific shear viscosity $\eta/s$ in terms of partonic differential cross section.
Note that an increasing running coupling $\alpha_s$ leads to smaller $\eta/s$. Equation (\ref{eq:etas})
represents a temperature-dependent estimate, as long as screen mass $\mu$ is not linear in
temperature~\cite{PhysRevC.92.014909}. In this work, we shall consider a constant screening mass, which results in
a rise of $\eta/s$ when temperature decreases. In the AMPT model, quarks and antiquarks combine to form
hadrons via a spatial coalescence model when scatterings stop.
The hadronic phase of the system evolves according to a relativistic transport model until hadrons freeze out.

\begin{figure}[b]
\begin{center}
\includegraphics[width=0.4\textwidth]{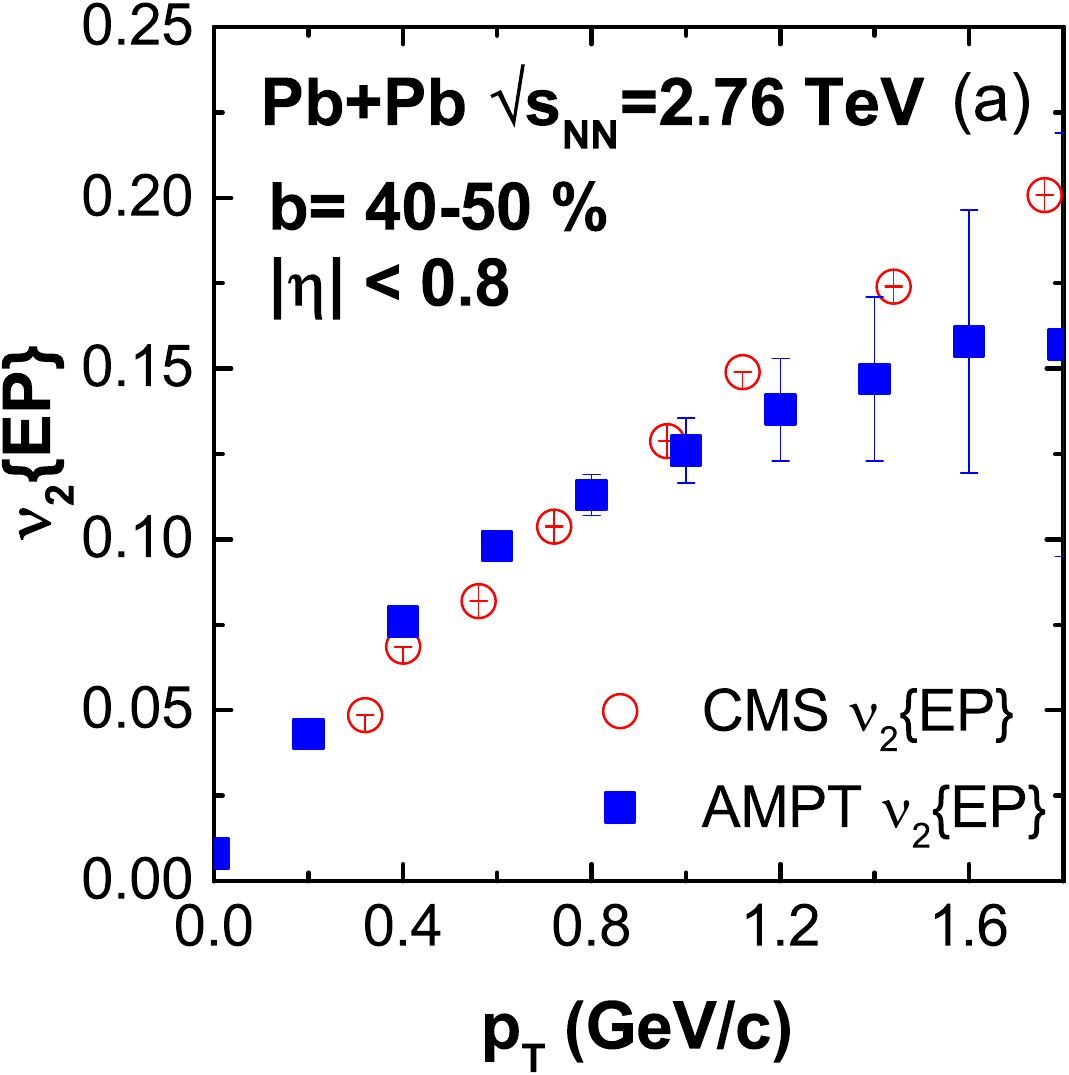}
\includegraphics[width=0.408\textwidth]{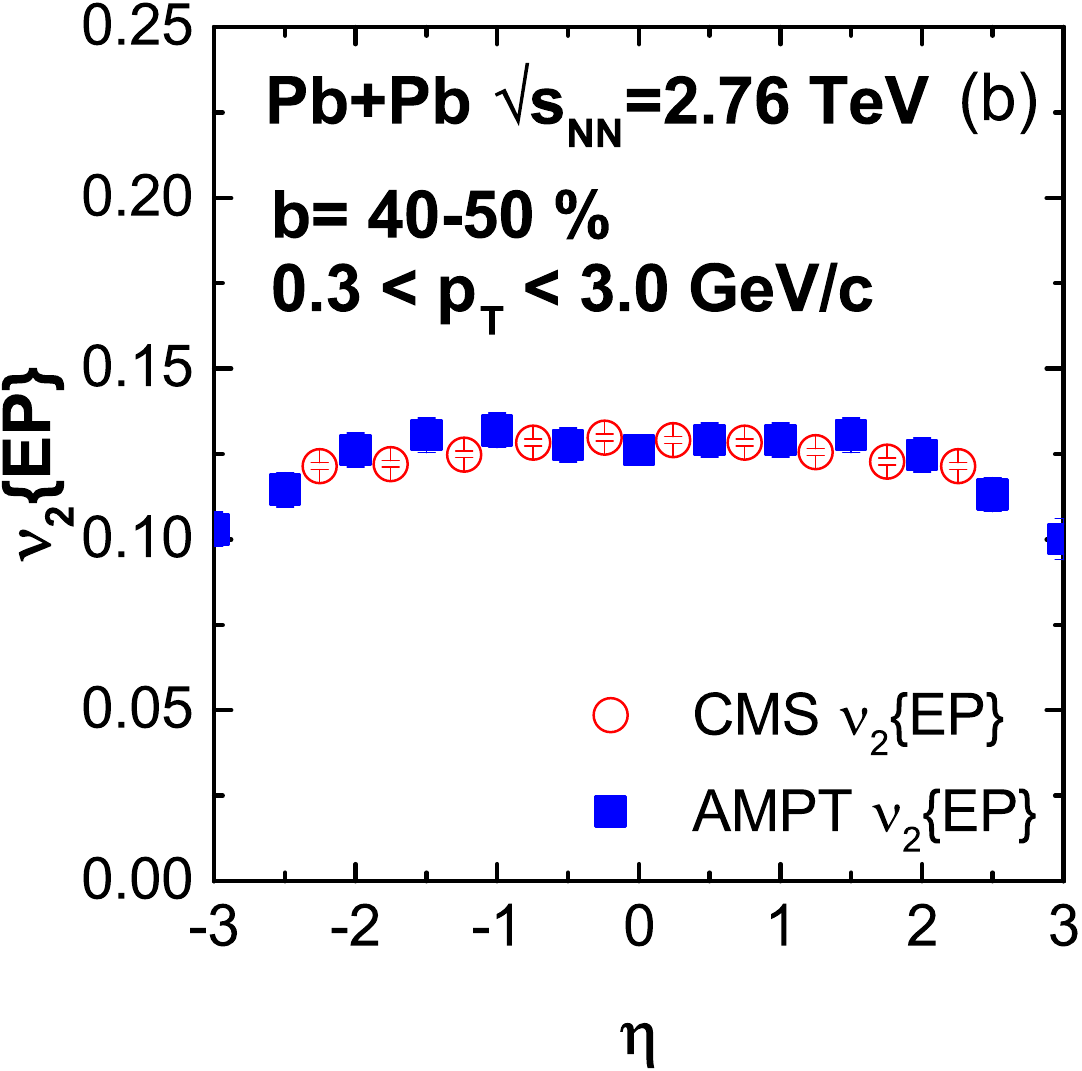}
\caption{(Color online) The (a) $p_{T}$- and (b) $\eta$- dependent elliptic flow $v_2$ measured
with respect to the event plane in the 40 - 50\% centrality class. Blue symbols are obtained from
AMPT simulations, in comparison with the CMS data at $\sqrt{s_{NN}}=2.76$ TeV~\cite{Chatrchyan:2012ta}.}
\label{fig1}
\end{center}
\end{figure}

Although in each single event only a finite number of particles are produced, the probability distribution of
these particles in azimuthal angle $f(\phi_p)$ can still be estimated, from which one can obtain
the complex flow harmonics $V_n$ using definition \Eq{eq:vn}. This complex $V_n$ in each event suffers
from statistical uncertainty due to finite number effect. A more systematic way to
calculate flow harmonics is to correlate particles from all events in one centrality class, as has been done in
experiments. From two-particle correlations, one obtains $v_n\{2\}$. From four-, six-, and
eight-particle correlations, one obtains higher order cumulants of flow harmonics: $v_n\{4\}$,
$v_n\{6\}$ and $v_n\{8\}$. An estimate of the event plane in the collisions can be done in a similar
manner, and correspondingly one has the flow harmonics measured with respect to
the event plane $v_n\{EP\}$.

By choosing appropriate parameters in the AMPT model according to \Ref{Xu:2011fi},
we are able to reproduce the measured observables at the LHC. For instance,
the differential
elliptic flow $v_2\{EP\}$ as a function of $p_T$ and pesudorapidity are shown in \Fig{fig1}
for the centrality class 40-50\% of Pb-Pb collisions at $\sqrt{s_{NN}}=2.76$ TeV, with
good agreements observed comparing to the CMS results.
In these calculations, in addition to parameters that control the Lund
string fragmentation, 
$a=0.5$ and $b=0.9$ GeV$^{-2}$,
$\alpha_{s}$ =0.33 and $\mu$ = 3.2 fm$^{-1}$ are chosen, so that effectively one has a
relatively large specific shear viscosity. At the initial temperature of LHC Pb-Pb collisions at $\sqrt{s_{NN}}=2.76$ TeV,
which is around $T\approx468$ MeV obtained by estimating the initial energy density, one
effectively has $\eta/s=0.273$ in the deconfined system.

\begin{figure}[b]
\begin{center}
\includegraphics[width=0.4\textwidth]{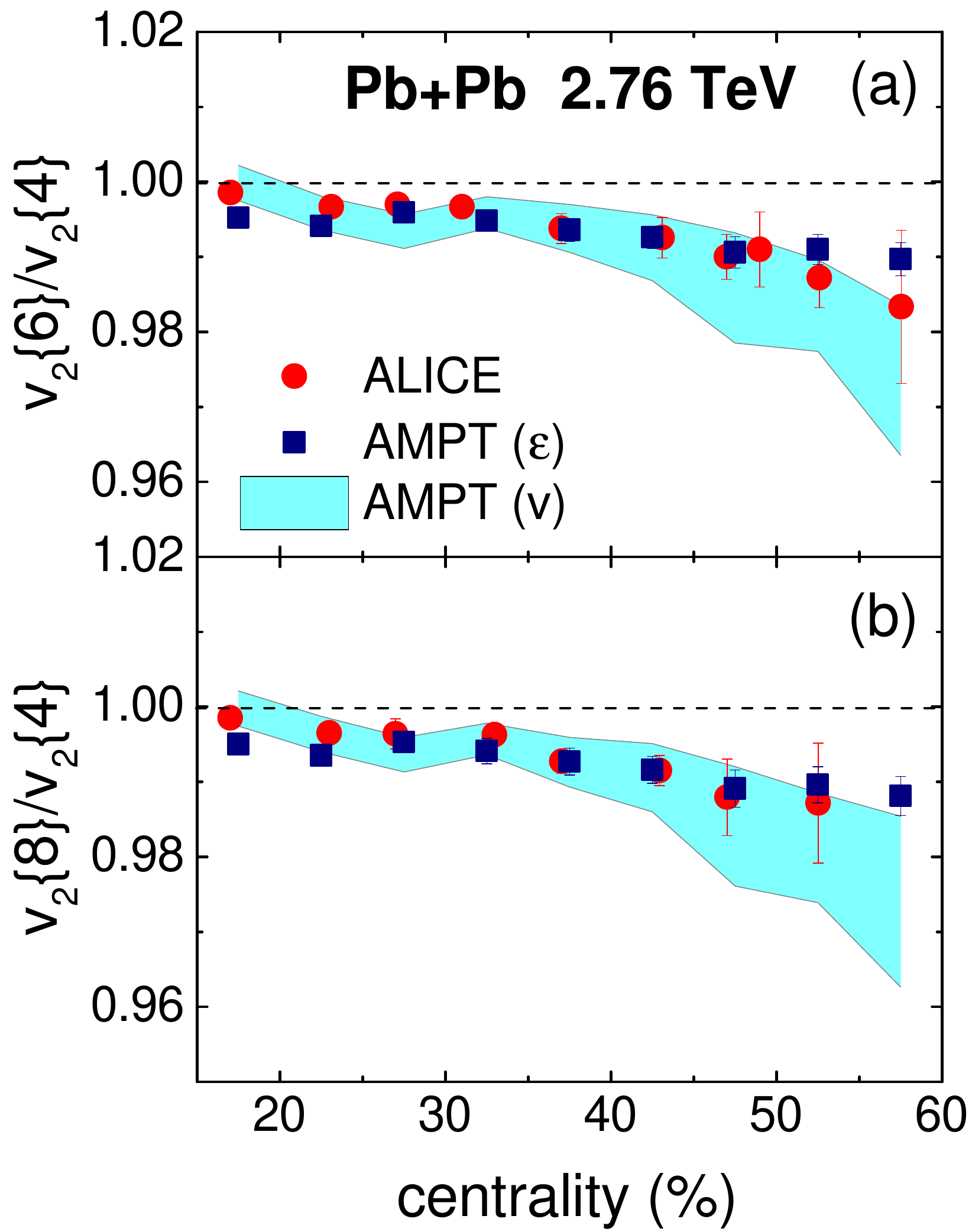}
\caption{(Color online)
Ratios of cumulants (a) $v_2\{6\}/v_2\{4\}$ and (b)
$v_2\{8\}/v_2\{4\}$.  Red points are from ALICE collaboration~\cite{Acharya:2018lmh}, colored bands are results
of AMPT calculations. The corresponding cumulant ratios of initial ellipticity $\varepsilon_2$ are shown
as blue squares.}
\label{fig2}
\end{center}
\end{figure}

In recent experiments at LHC energies, more sophisticated measurements of elliptic flow
have been carried out, revealing the fluctuating nature of $v_2$ (cf~\Ref{Adam:2016izf,Sirunyan:2017fts,Acharya:2018lmh}). 
To estimate the fluctuation effect of $v_2$,
for the Pb-Pb collisions at $\sqrt{s_{NN}}=2.76$ TeV, in each 5\%-centrality bin from the 15\% to
60\%, we generate approximately 5000 events from AMPT simulations.
In \Fig{fig2}, our AMPT results for the ratio of $v_2\{6\}/v_2\{4\}$ and ratio
$v_2\{8\}/v_2\{4\}$ are presented as colored bands
as a function of centrality percentile.
The width of bands corresponds to statistical errors due to finite number effect,
which in our calculations are estimated via a jackknife resampling.
In comparison to the experiments from the ALICE Collaboration (red points), an overall
agreement is observed within errors, which indicates that AMPT model is able to
capture the non-Gaussian properties of $v_2$ fluctuations.

One should be aware that both ratios are less than unity, and also
the fact that $v_2\{8\}$ is smaller than $v_2\{6\}$, are characteristic natures of flow fluctuations
known from hydrodynamic modeling. In hydrodynamic modeling of heavy-ion collisions, fluctuations
of elliptic flow $v_2$ are mostly determined by fluctuations of initial ellipticity $\varepsilon_2$, due to
the fact that $V_2\propto\E_2$. This is also observed in our AMPT calculations. As shown
in \Fig{fig2} the ratio of cumulants of $\varepsilon_2$ (blue squares) are compatible with those
of $v_2$, except for very peripheral collisions where the ratios of $\varepsilon_2$ are slightly larger.

\begin{figure}
\begin{center}
\includegraphics[width=0.4\textwidth]{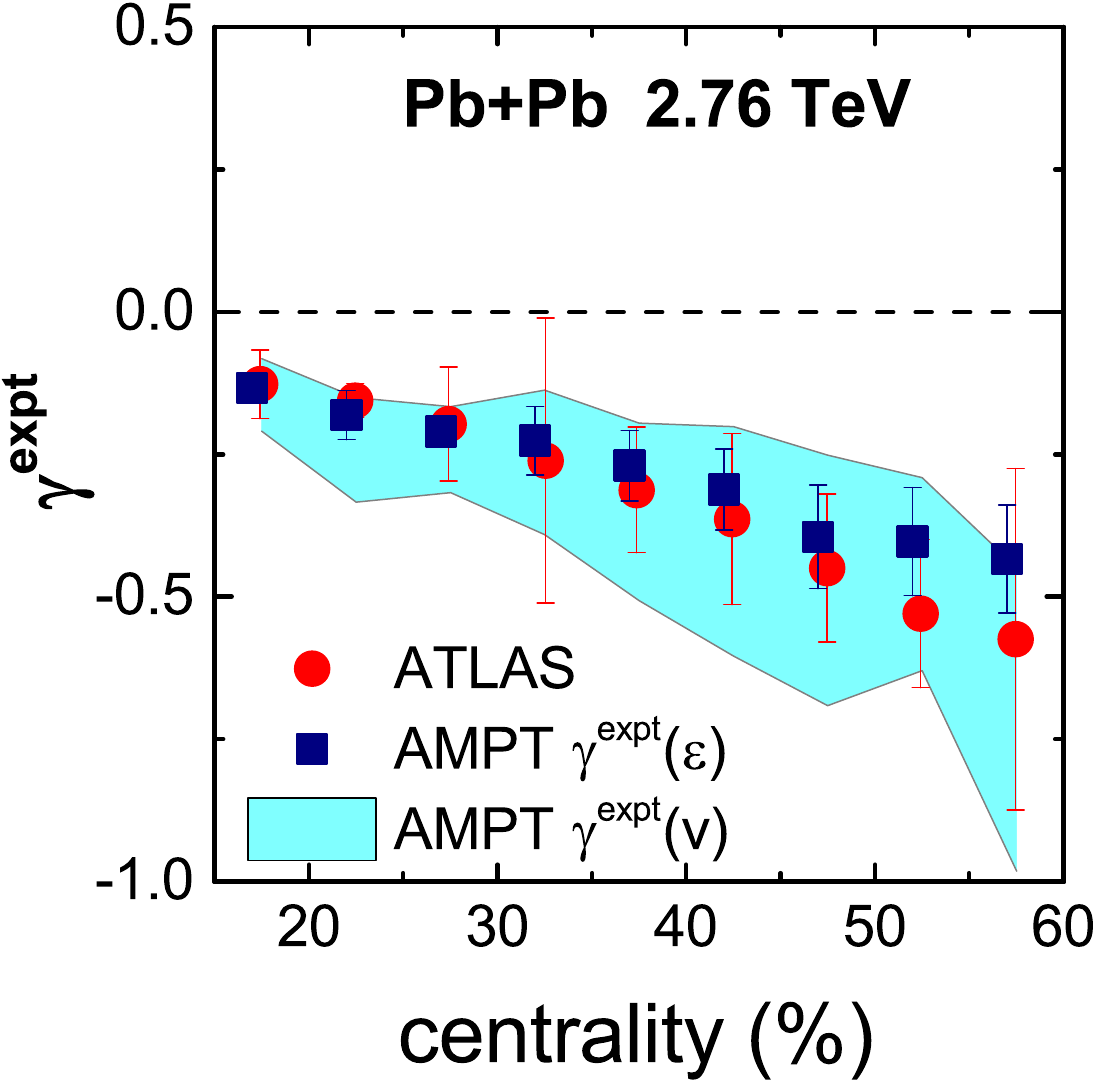}
\caption{(Color online)
Standardized skewness of $v_2$ fluctuations in Pb-Pb collisions as a function of centrality from
AMPT simulations (colored band) and ATLAS~\cite{Giacalone:2016eyu} data
(red points). The corresponding results of standardized skewness of initial state $\varepsilon_2$
are shown as blue squares. }
\label{fig3}
\end{center}
\end{figure}

The information of $v_2$ fluctuations can be as well captured by skewness. Given $v_2\{2\}$, $v_2\{4\}$ and
$v_2\{6\}$, one may estimate the standardized skewness as~\cite{Giacalone:2016eyu}
\be
\gamma^{expt} \equiv -6\sqrt{2}v_{2}\{4\}^{2}\frac{v_{2}\{4\}-v_{2}\{6\}}{[v_{2}\{2\}^{2}-v_{2}\{4\}^{2}]^{3/2}}.
\ee
\Fig{fig3} depicts the estimated standardized skewness of $v_2$ fluctuations from our AMPT
simulations, as colored bands. Again, the width of a band is determined by statistical errors via a
jackknife resampling.
AMPT results agree well with the recent experimental data (red points). One observes a negative
value of the skewness, with its magnitude increasing as centrality percentile grows.
In hydrodynamic modeling, this negative skewness of $v_2$ fluctuations is understood as
a consequence of the negative skewness of $\varepsilon_2$, due to the combined effect of
an upper bound $\varepsilon_2<1$ and a nonzero mean of ellipticity in the reaction plane.
Similarly, we notice that, in the case of AMPT simulations, skewness of $v_2$ is comparable with
that of initial $\varepsilon_2$.

\section{HYDRODYNAMIC RESPONSE RELATION IN AMPT}
\label{sec:sec3}

\begin{figure*}[!t]
\begin{center}
\includegraphics[width=0.4\textwidth]{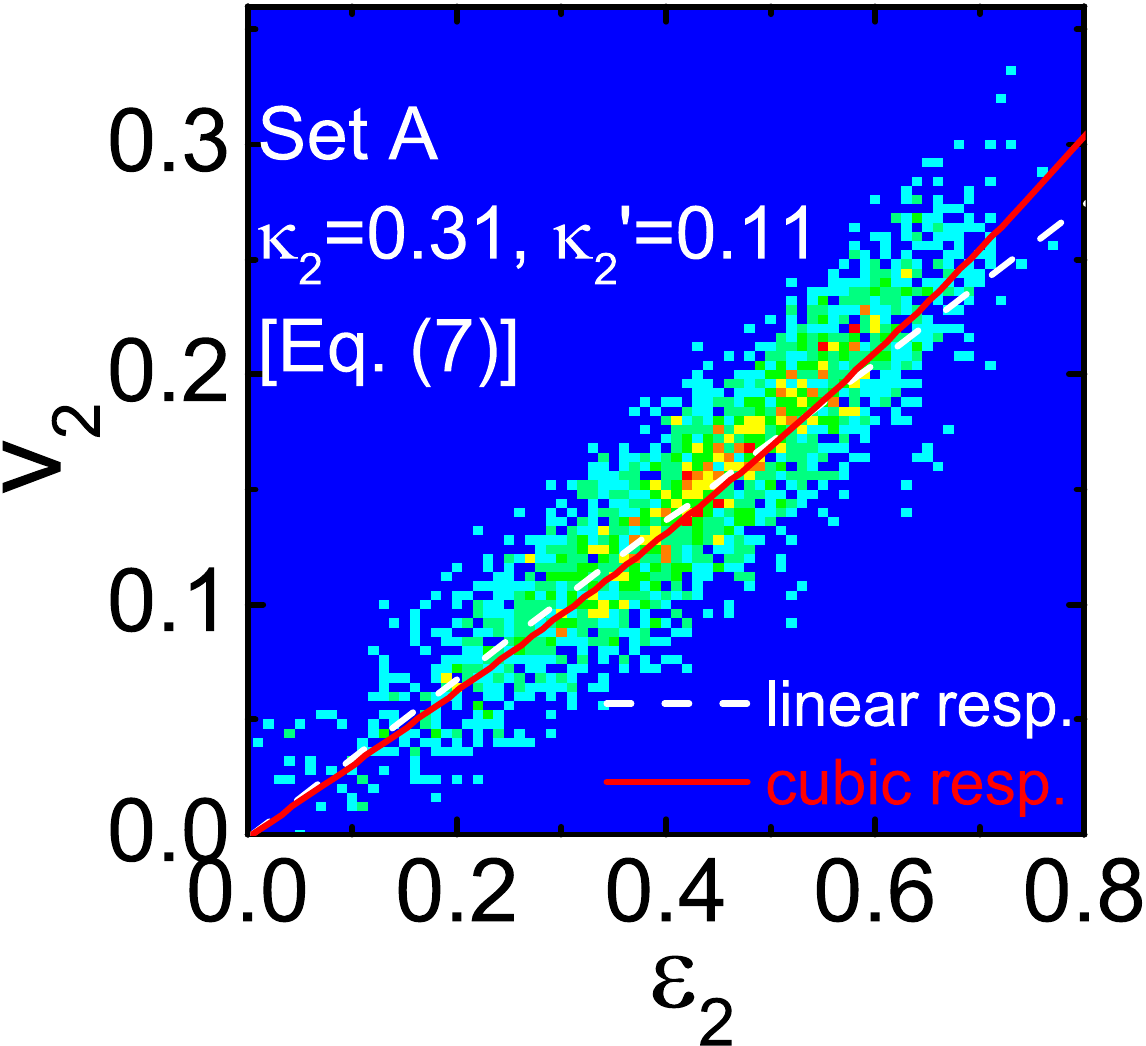}
\includegraphics[width=0.4\textwidth]{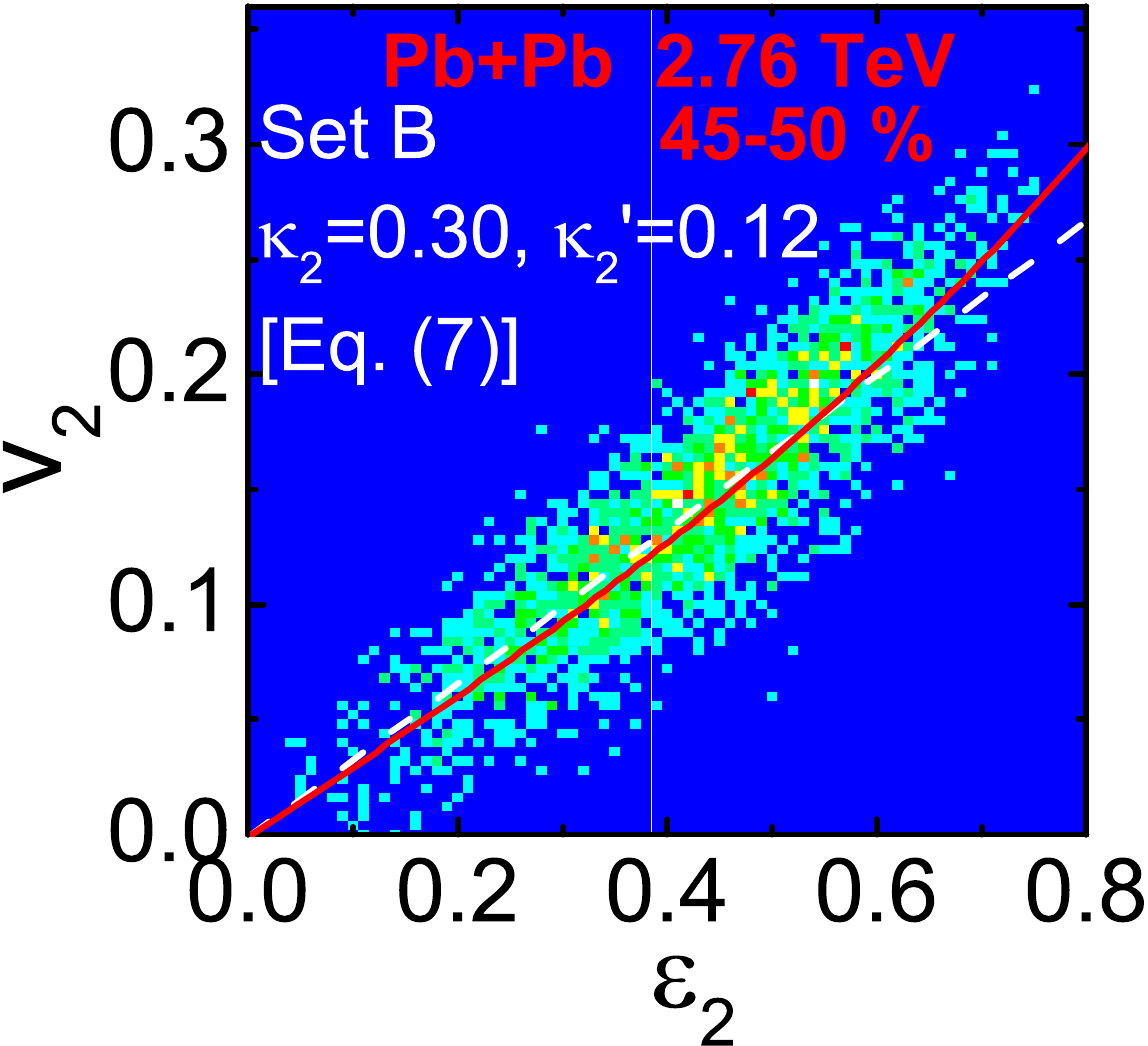} \\
\includegraphics[width=0.4\textwidth]{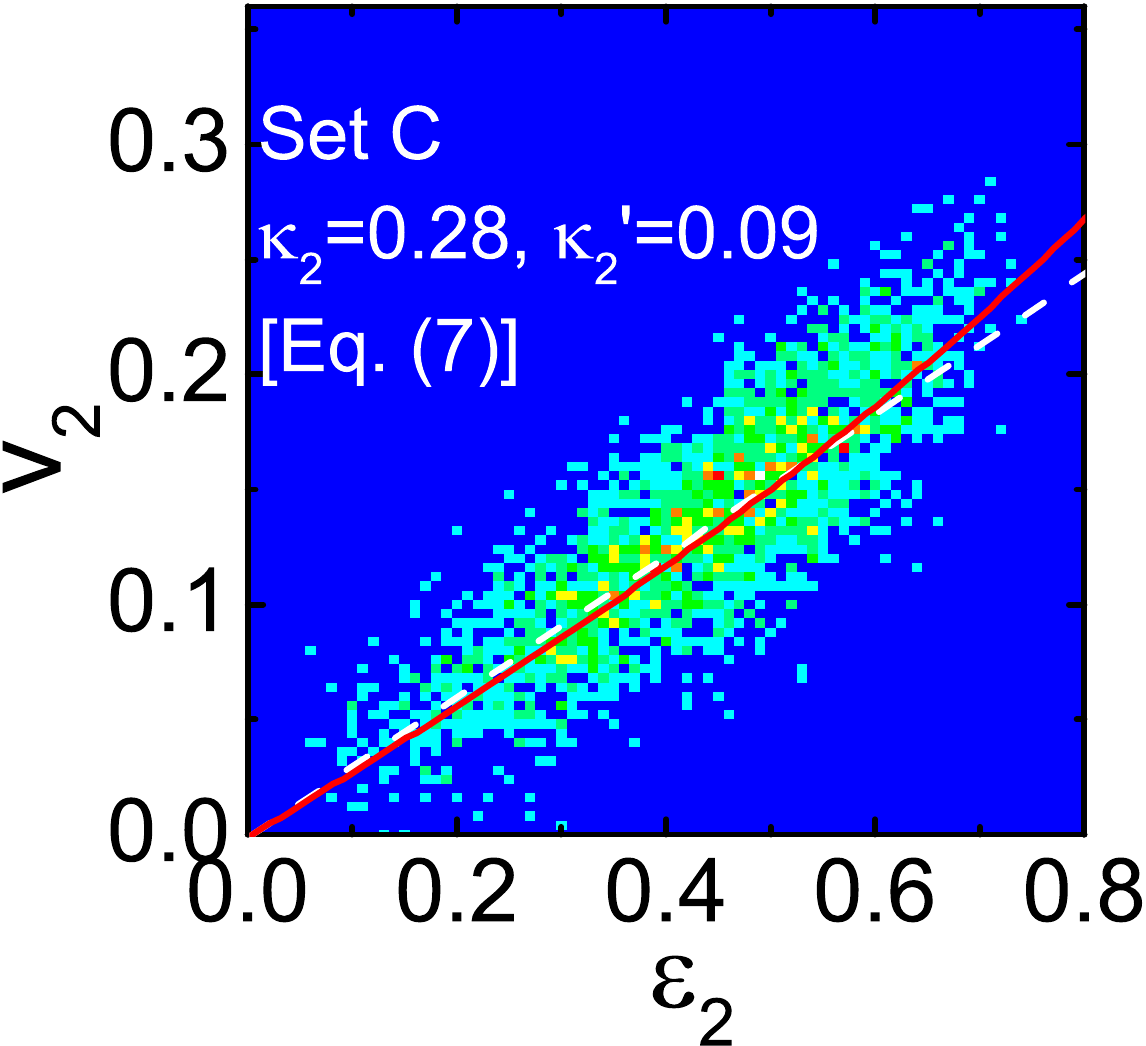}
\includegraphics[width=0.4\textwidth]{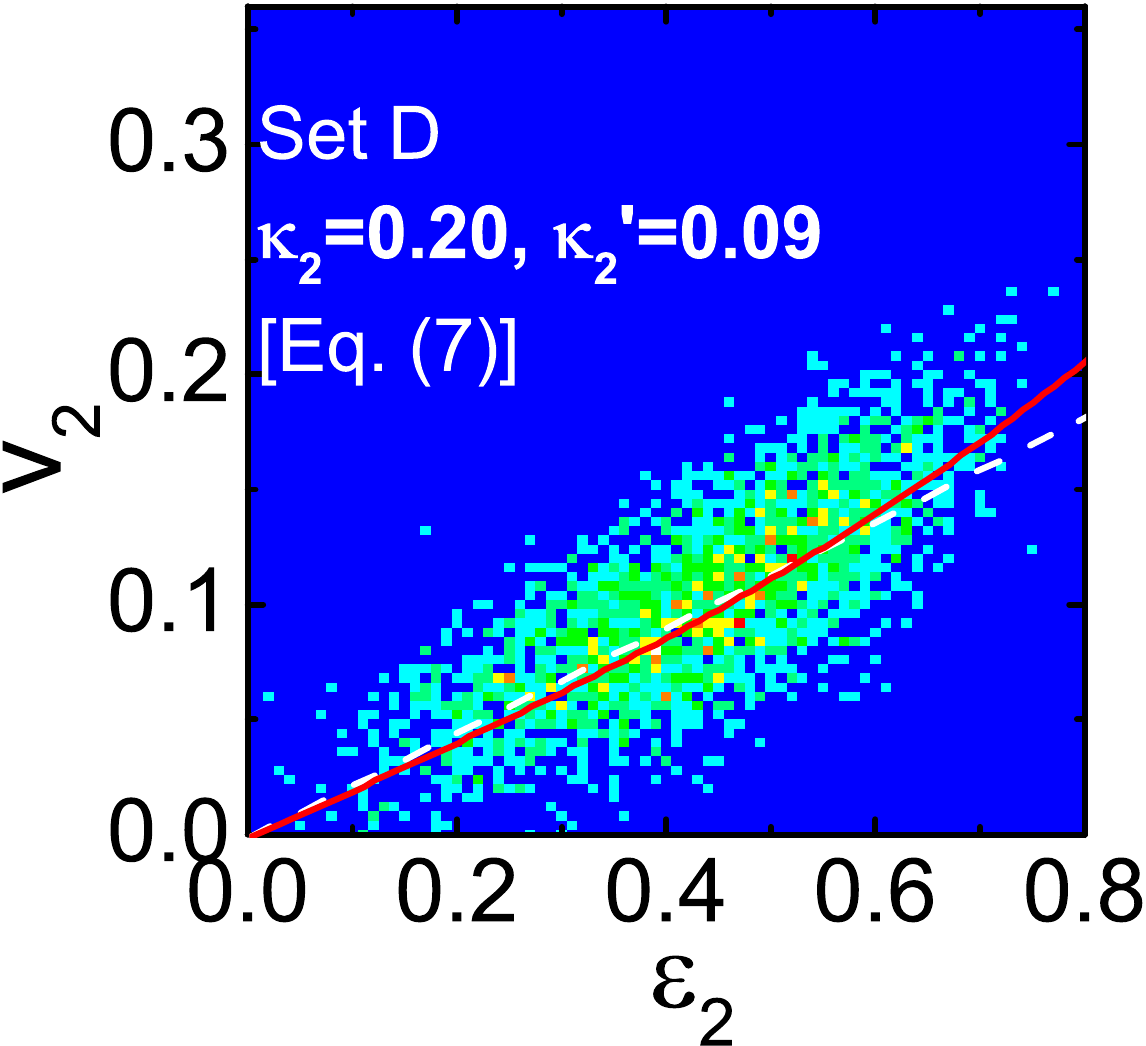}
\caption{(Color online)
Scatter plot of event-by-event $v_2$ from AMPT simulations for Pb-Pb
collisions, as a function of $\varepsilon_2$. We take model parameters so that the effective
specific shear viscosity $\eta/s$ is increased from (a) 0.08, to (b) 0.10, (c) 0.14, and (d) 0.273,
at temperature $T=468$ MeV.
Red solid lines are from relation \Eq{eq:resp1} with a cubic-order correction, while
white dashed lines correspond to linear response relation.
}
\label{fig4}
\end{center}
\end{figure*}

In the previous section we have seen $v_2$ fluctuations from
AMPT simulations, in terms of the cumulant of $v_2$ from multiparticle
correlations and standardized skewness. Especially, the fluctuations of $v_2$ follow
to a large extent the fluctuations of $\varepsilon_2$.  This feature is very similar
to what one would expect in a hydrodynamic modeling of heavy-ion collisions, in which
a hydrodynamic response relation between $V_2$ and $\E_2$ has been
established~\cite{Noronha-Hostler:2015dbi},
\be
\label{eq:resp}
V_2=\kappa_2 \E_2 + \kappa_2' \varepsilon_2^2\E_2 + \delta_2
\ee
Equation~(\ref{eq:resp}) is achieved by regarding initial state eccentricity $\varepsilon_n$ as small
quantities, hence one may expand the complex quantities $V_2$ in terms of $\E_n$.
Owing to the condition of rotational symmetry, the leading-order term is a linear response
proportional to
$\E_2$, with $\kappa_2$ the linear response coefficient determined by medium dynamical
expansion.  The next-leading-order contribution is of cubic order, and is dominantly determined
by $O(\varepsilon_2^3)$. Apparently, the cubic-order contribution is not important unless
$\varepsilon_2$ becomes large, as in peripheral collisions. The quantity
$\delta_2$ in \Eq{eq:resp} describes additional event-by-event fluctuations, which affects the
response relation on an event-by-event basis.

Although $V_2$ fluctuates
from event to event, as well as $\E_2$, both the linear response
coefficient $\kappa_2$ and cubic-order response coefficient $\kappa_2'$ are
considered constant in each centrality class. By minimizing the effect of additional
fluctuations $\delta_2$, one solves $\kappa_2$ and $\kappa_2'$~\cite{Noronha-Hostler:2015dbi},
\begin{subequations}
\label{kappaprime}
\begin{align}
\kappa_2&=\frac{
{\rm Re}\left(
\langle \varepsilon_2^6\rangle\langle  V_2\E_2^*\rangle
-\langle \varepsilon_2^4\rangle\langle
V_2\E_2^*\varepsilon_2^2\rangle\right)}
{\langle \varepsilon_2^6\rangle\langle \varepsilon_2^2\rangle-\langle
 \varepsilon_2^4\rangle^2}\\
\kappa'_2&=\frac{
{\rm Re}\left(
-\langle \varepsilon_2^4\rangle\langle  V_2\E_2^*\rangle
+\langle \varepsilon_2^2\rangle\langle
V_2\E_2^*|\varepsilon_2|^2\rangle\right)}
{\langle \varepsilon_2^6\rangle\langle \varepsilon_2^2\rangle-\langle
  \varepsilon_2^4\rangle^2},
\end{align}
\end{subequations}
where bracket $\bra \ldots\ket$ indicates average over events. Note that
if one ignores contribution from the cubic-order response, $\kappa_2$ in \Eq{kappaprime}
 reduces to
 \be
 \label{eq:resp0}
 \kappa_2 = \frac{{\rm Re}\bra V_2\E_2^*\ket}{\bra \varepsilon_2^2\ket}
 \ee
 One may check that a cubic-order correction reduces slightly
 the linear response coefficient, comparing \Eq{eq:resp0} to \Eq{kappaprime}.

Equation~(\ref{eq:resp}) in hydrodynamic modeling has been verified by event-by-event hydrodynamic
simulations~\cite{Noronha-Hostler:2015dbi,Niemi:2015voa,Gardim:2011xv}.
It is interesting to test the response relation in AMPT model.
In the present setup of AMPT simulations for Pb-Pb collisions at $\sqrt{s_{NN}}=2.76$ TeV,
we focus on the centrality class 45-50\%. In hydrodynamic modeling, in
45-50\% centrality class, both linear and nonlinear response are found important.
We generate approximately
5000 events in our AMPT simulations. A scatter plot of $v_2$ versus $\varepsilon_2$ is
obtained and is shown \Fig{fig4} (d). Each point in \Fig{fig4} (d) corresponds to one
collision event. It is worth mentioning that the statistical uncertainty of $v_2$ in
each event due to finite multiplicity is not included, which would in principle lead to a smearing
along $v_2$ in \Fig{fig4}. In \Fig{fig4} (d),
these points distribute along a line, except for a slight tilde
at large values of $\varepsilon_2$ which implies nonlinearity.
We find that a response relation between magnitudes
derived from \Eq{eq:resp}, describes well the trend,
\be
\label{eq:resp1}
v_2=\kappa_2\varepsilon_2 + \kappa'_2 \varepsilon_2^3,
\ee
similar to what one would expect from hydrodynamics.
Given these solved values of $\kappa_2$ and $\kappa_2'$ according to
\Eq{kappaprime}, \Eq{eq:resp1} is plotted in \Fig{fig4} (d) as the red solid line.
Without the cubic-order correction, one has $v_2=\kappa_2\varepsilon_2$ which
is shown as white dashed line in the figure.  It should be emphasized
that the red line and white dashed line
are \emph{not} fitting the scattering points, but were determined with respect to the solved values
of $\kappa_2$ and $\kappa_2'$ according to \Eq{kappaprime}.
Note also that the resulting linear response coefficient
$\kappa_2$ are not identical with or without cubic-order corrections.
Dispersion around the linear and cubic-order response reflects event-by-event fluctuations.
The width of the dispersion is related to fluctuation strength.

\subsection{Effect of $\eta/s$}

In hydrodynamics, the effects of $\eta/s$
are twofold. First, it crucially determines medium response,
\ie, $\kappa_2$ and $\kappa_2'$. When $\eta/s$ increases,
the linear response coefficient is suppressed.
Second, from hydrodynamic simulations, it has also been noticed that
event-by-event fluctuations around hydrodynamic response relations are reduced with respect to
a larger value of $\eta/s$~\cite{Noronha-Hostler:2015dbi,Niemi:2015voa}.

To study the dissipative effect in AMPT simulations,
by changing parameters for the parton cascade and for the Lund string fragmentation, we
adjust effectively the ratio of shear viscosity to entropy density $\eta/s$
of partons, with respect to \Eq{eq:etas}.
We simulate using the AMPT model in the same
centrality class (45-50\%),  keeping total multiplicity unchanged but
varying $\eta/s$. We summarize these four sets of parameters used in AMPT simulations
in \Tab{tab1}.
In addition to set D that has been used in previous sections
to describe the $\sqrt{s_{NN}}=2.76$ TeV Pb-Pb collisions,
which has, at $T=468$ MeV, $\eta/s=0.273$,
parameter set A, B, and C lead to, at $T=468$ MeV, $\eta/s=0.08$, $0.10$ and $0.14$
respectively.
Correspondingly, results with respect to these simulations are shown in \Fig{fig4} (a),
\Fig{fig4} (b) and \Fig{fig4} (c).

\begin{table}
\caption{ Parameters used in AMPT simulations corresponding to
different values of $\eta/s$ at $T=468$ MeV. Parameters
are taken according to \Ref{Solanki:2012ne,Xu:2011fe}.
\label{tab1}
}
\setlength{\tabcolsep}{3.3mm}{
\begin{tabular}{l c c c r}
 \hline\hline
 & & & Set &  \\ \cline{2-5}
 & A  & B & C & D \\ \hline
 $\eta/s$ ($T=468$ MeV) & 0.08 &  0.10 & 0.14 & 0.273 \\ 
 $a$ & 2.2 &  2.2 & 2.2 & 0.5 \\
 $b$ (GeV$^{-2})$ & 0.5 &  0.5 & 0.5 & 0.9 \\
 $\alpha_s$ & 0.47 & 0.47 & 0.47 & 0.33 \\
 $\mu$ (fm$^{-1})$ & 1.8 & 2.3 & 3.2 & 3.2 \\
 \hline
\end{tabular}
}
\end{table}

\begin{figure}[b]
\begin{center}
\includegraphics[width=0.4\textwidth]{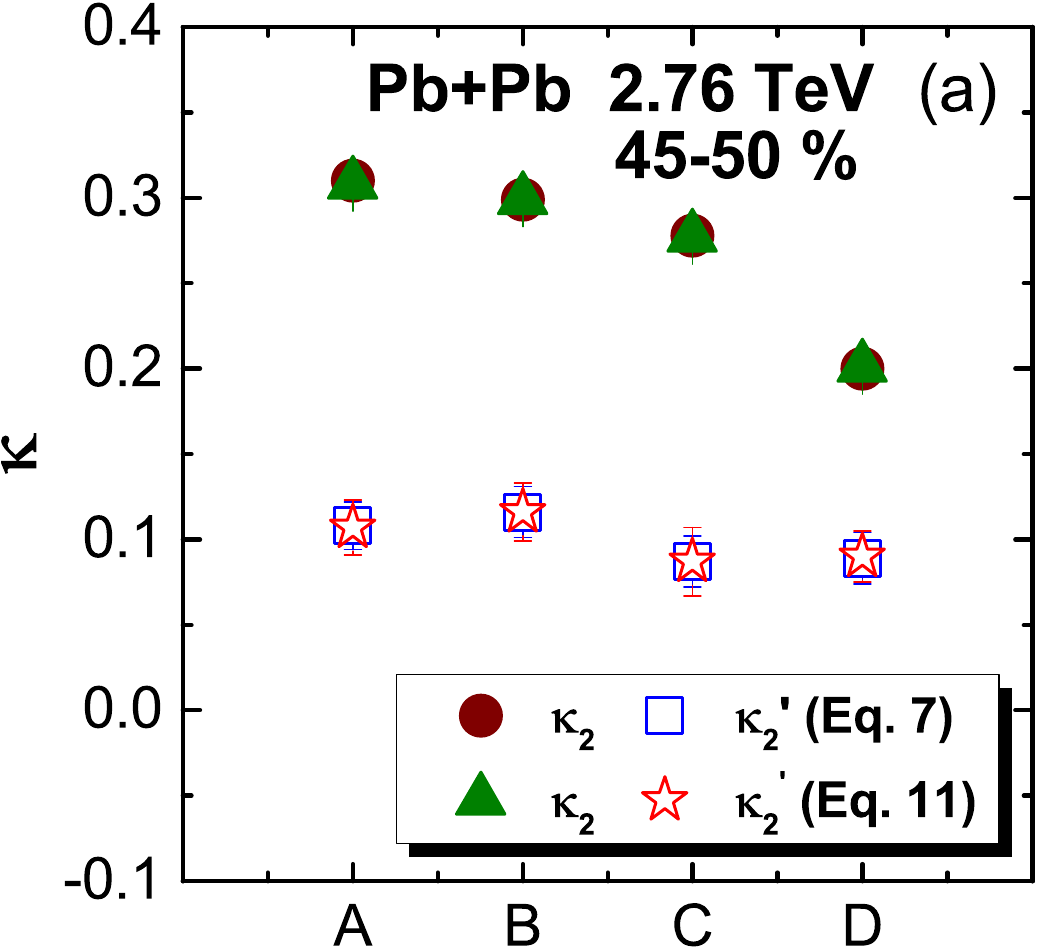}
\includegraphics[width=0.4\textwidth]{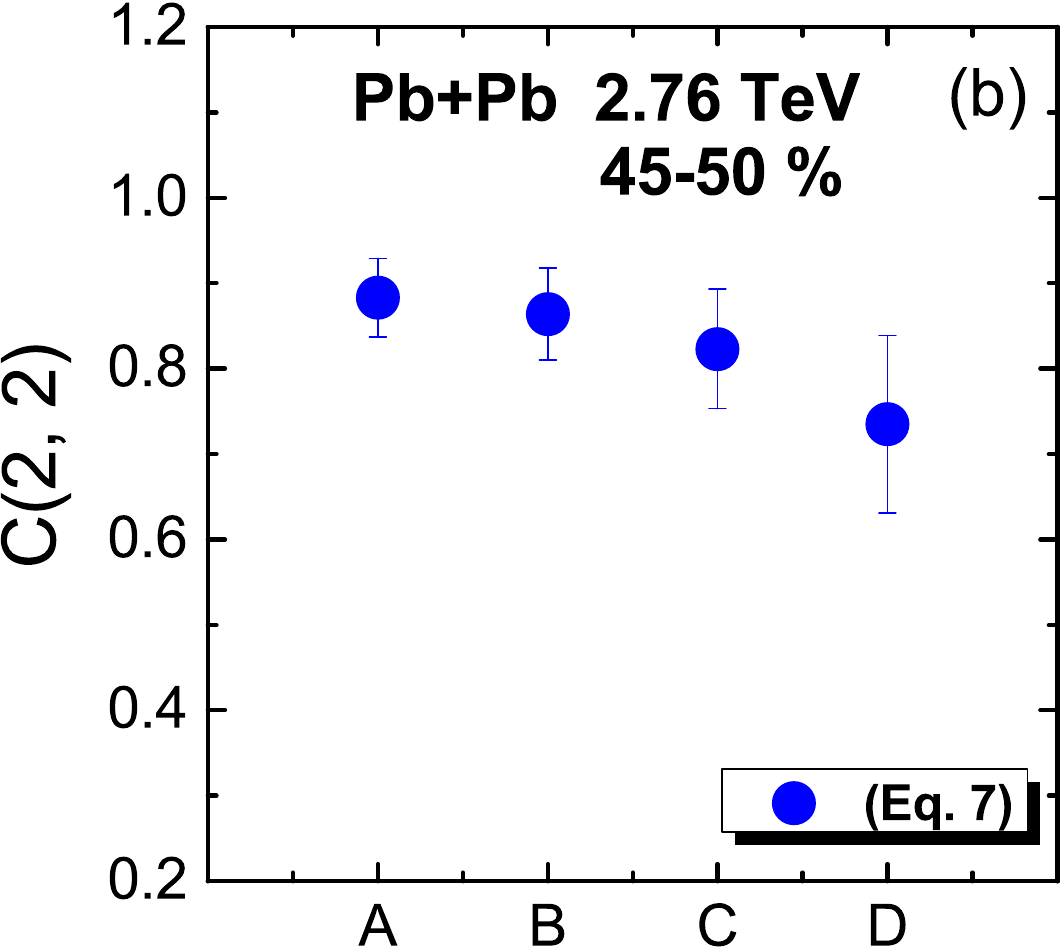}
\caption{(Color online) (a) AMPT results of the linear response and cubic response coefficients of
the Pb-Pb 45-50\% centrality class for four different sets of parameters corresponding to an increase of
$\eta/s$ from set A to set D, 
with respect
to \Eq{kappaprime} and \Eq{eq:nonflow_v2} with nonflow subtraction.
(b) Pearson correlation coefficient $C(2,2)$ for different sets of parameters.
}
\label{fig5}
\end{center}
\end{figure}

For each set of $\eta/s$, we calculate
linear and cubic response coefficients, $\kappa_2$ and $\kappa_2'$, giving rise
to the response relation with (red solid lines) or without (white dashed lines)
cubic-order corrections. As anticipated, the value of linear response coefficient, the slope of lines
in \Fig{fig4}, is reduces as $\eta/s$ increases, consistency with hydrodynamic modeling of heavy-ion
collisions. We plot in \Fig{fig5} (a) the obtained value of linear and cubic response coefficients.
The cubic-order response coefficient
remains approximately constant, $\kappa_2'\approx0.1$. 

In contrast to hydrodynamic modeling,
the event-by-event fluctuations are not suppressed by viscosity in the AMPT
simulations, as
can be seen from the width of dispersion in \Fig{fig4}. In fact, one finds an slight increase of the
fluctuation strength, if it is measured relative to the mean of $v_2$,
$\bra |\delta_2|^2\ket/\bra v_2^2\ket$. 
To quantify these effects,
we measure the correlation of complex variables $V_2$ and $\E_2$, by the Pearson
correlation coefficient,
\be
C(2,2)=\frac{\rm{Re}\bra V_2 \E_2^*\ket}{\sqrt{\bra v_2^2\ket\bra\varepsilon_2^2\ket}}\,.
\ee
The Pearson correlation coefficient $C(2,2)$ captures simultaneously correlation between
magnitudes and phases.
An absolute correlation is approached when $C(2,2)=1$, while $C(2,2)=0$ indicates
no correlation. Since fluctuations tend to break correlation between $V_2$ and $\E_2$,
the effect of fluctuation reduces $C(2,2)$. In \Fig{fig5} (b), we indeed find that $C(2,2)$ decreases
linearly as $\eta/s$ increases from set A to set D.

\subsection{Nonflow subtraction}

We have studied the linear and cubic response which relates elliptic flow $V_2$ and
initial ellipticity $\E_2$ via AMPT simulations. Although the strategy is very similar
to hydrodynamic simulations, with the linear and cubic-order response coefficients obtained through
minimizing event-by-event fluctuations,
AMPT simulations contain nonflow effects. These nonflow
effects are beyond pure hydrodynamic calculations, including, \emph{e.g.}, short-ranged correlations in
the deconfined medium from particle scatterings, hence they are nonhydrodynamic.
Nonetheless, considering the fact that linear and cubic response relations are hydrodynamic and
are dominated by the evolution of long-wavelength modes of the medium system, one
would expect that the response relations depend little on nonflow effects.

\begin{figure}[b]
\begin{center}
\includegraphics[width=0.4\textwidth]{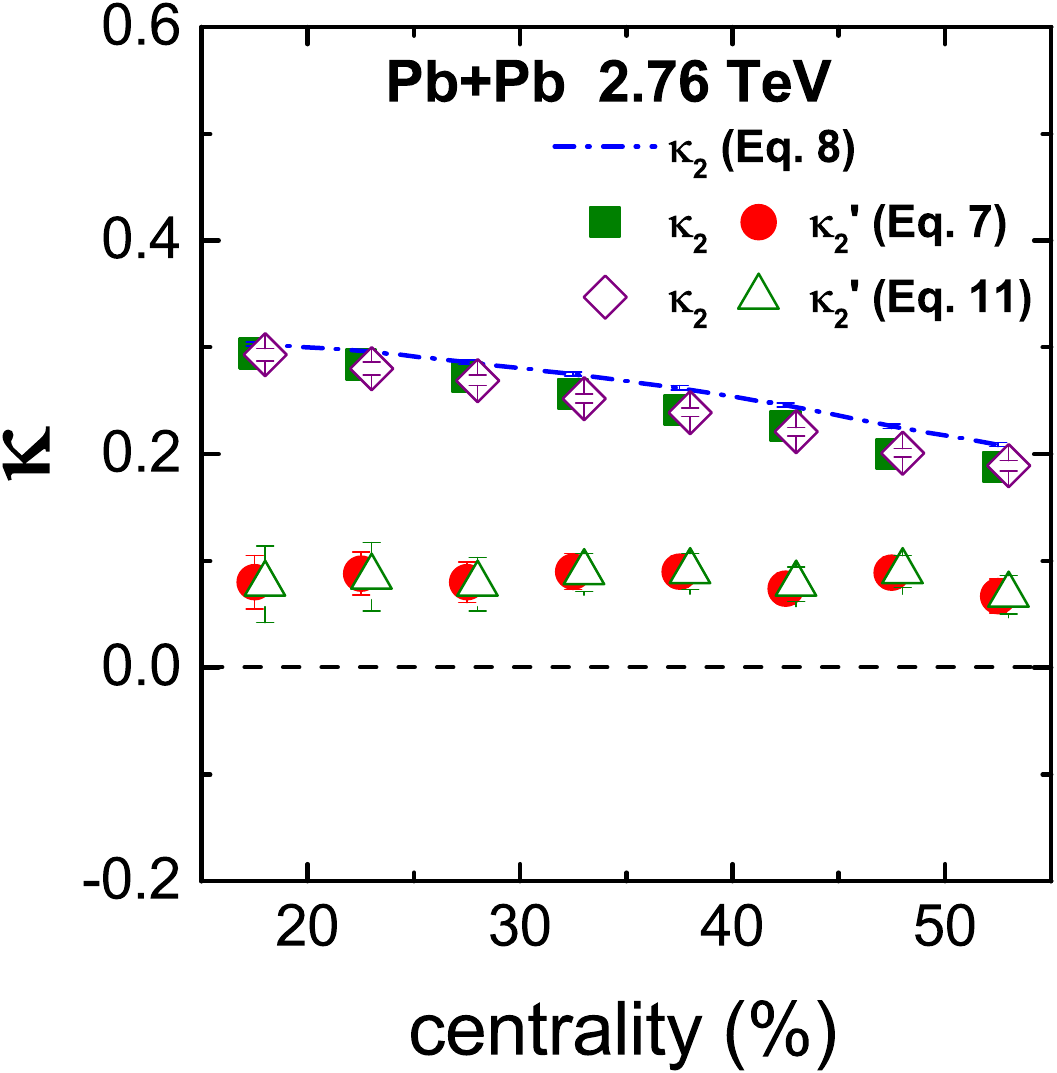}
\caption{(Color online) Linear and cubic-order response coefficients $\kappa_2$ and
$\kappa_2'$ as a function of centrality percentile, from AMPT simulations of Pb-Pb
collisions at $\sqrt{s_{NN}}=2.76$ TeV. The results of
$\kappa_2$ and $\kappa_2'$ are calculated according to \Eq{kappaprime} and \Eq{eq:nonflow_v2}
corresponding to nonflow effects subtraction, respectively. For comparison, the linear
response coefficient $\kappa$ calculated with respect to \Eq{eq:resp0} is
shown as dashed line.
}
\label{fig6}
\end{center}
\end{figure}

In order to test the nonflow effects on the linear and cubic response,
we subtract nonflow contributions in the AMPT simulations.
To subtract nonflow effects in the flow harmonics, one may either
take a pseudorapidity gap or rely on multiparticle cumulants. Both methods have been
applied extensively in experiments~\cite{Chatrchyan:2013kba,Acharya:2018zuq,Aad:2014vba}.
Therefore, based on the response relation in \Eq{eq:resp},
we find from the two-particle and four-particle correlations,
    \begin{subequations}
    \label{eq:nonflow_v2}
    \begin{align}
    v_{2}\{2,|\Delta\eta|\} &= \kappa_{2}\varepsilon_{2}\{2\}\left[1 + \frac{\kappa_{2}^{'}}{\kappa_{2}}\frac{\langle\varepsilon_{2}^{4}\rangle}{\langle\varepsilon_{2}^{2}\rangle}\right]\,, \\
    v_{2}\{4\} &= \kappa_{2}\varepsilon_{2}\{4\}\left[1 + \frac{\kappa_{2}^{'}}{\kappa_{2}}\frac{2\langle\varepsilon_{2}^{2}\rangle\langle\varepsilon_{2}^{4}\rangle-\langle\varepsilon_{2}^{6}\rangle}{2\langle\varepsilon_{2}^{2}\rangle^{2}-\langle\varepsilon_{2}^{4}\rangle}\right]\,.
    \end{align}
    \end{subequations}
In writing \Eq{eq:nonflow_v2}, we have assumed that a pseudo-rapidity gap is sufficient to
take out the nonflow contribution, \ie, $\delta_2$, in $v_2\{2\}$. In practice, a
 pseudorapidity gap $|\Delta \eta|>1$ is taken into accout in our AMPT simulations for $v_{2}\{2,|\Delta\eta|\}$. Similarly,
$\delta_2$ does not appear in $v_2\{4\}$. Equation~(\ref{eq:nonflow_v2}) then allows us to solve $\kappa_2$
and $\kappa_2'$, without nonflow effects. The corresponding
results of response coefficients are shown in \Fig{fig5} (a) as a function of
$\eta/s$, and in \Fig{fig6} as a function of centrality percentile.
As expected, numerical solutions of the
response coefficients are found to be compatible with or without nonflow contributions.

\section{SUMMARY AND DISCUSSIONS}

In this work, we have carried out AMPT simulations for Pb-Pb collisions at the
LHC energy $\sqrt{s_{NN}}=2.76$ TeV.
Our AMPT results of ellitpic flow, especially the flucuations of $v_2$ are
compatible with experiments. In addition, we found that the flucutation behavior of
$v_2$, characterized in terms of cumulant ratios or the standardized skewness, is
closely related to that of $\varepsilon_2$. This feature has been noticed in
hydrodynamic modelings, where the hydrodynamic response relations were proposed to explain the generation
of harmonic flow. In the AMPT model, we observed very similar response relations, which
can be well described by proper linear and cubic-order response coefficients.
This observation confirms the fact that
elliptic flow $v_2$ is indeed a consequence of medium response to initial state $\varepsilon_2$.
Since the medium
response reflects long-wavelength mode evolution, which is hydrodynamic,
similarity to what has been found in hydrodynamic modelings is understandable.
However, AMPT simulations contain extra event-by-event fluctuations due to nonflow effects. Even though
these fluctuations are nonhydrodynamic, as they are not sensitive to dissipative effect of the
medium system, they do not affect the response relations.

\section*{ACKNOWLEDGMENTS}
We thank Jean-Yves Ollitrault and Zi-Wei Lin for very helpful discussions. D.X.W. and X.G.H. are supported by the Young 1000 Talents Program of China, NSFC with Grant No.~11535012 and No.~11675041. L.Y. is supported in part by the
Natural Sciences and Engineering Research Council of Canada.

\appendix
\section{AMPT RESULTS OF Pb-Pb COLLISIONS AT $\sqrt{s_{NN}}=5.02$ TeV}
\label{app1}

\begin{figure}
\begin{center}
\includegraphics[width=0.4\textwidth]{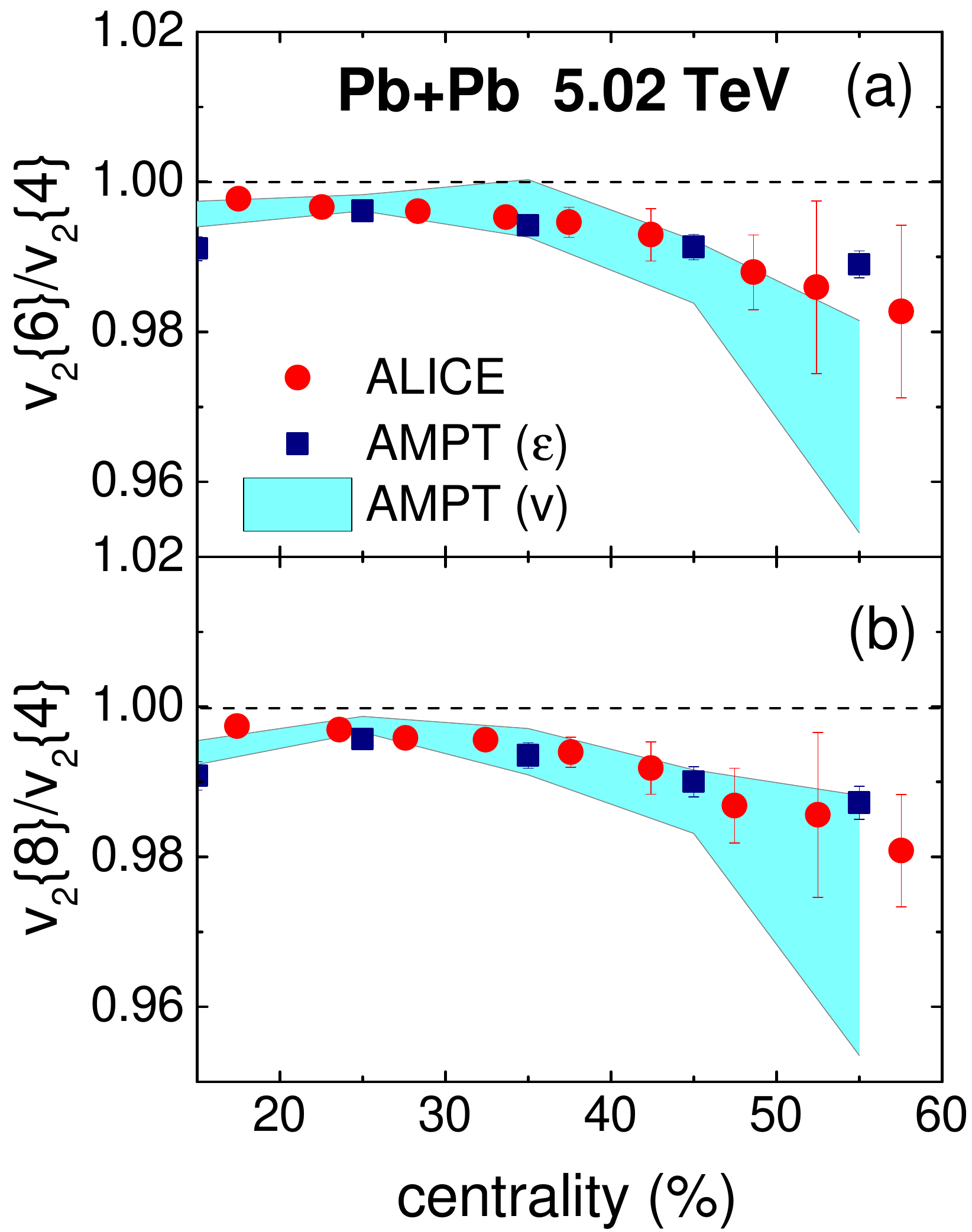}
\caption{(Color online)
Ratios of cumulants (a) $v_2\{6\}/v_2\{4\}$ and (b)
$v_2\{8\}/v_2\{4\}$ for Pb-Pb collisions at $\sqrt{s_{NN}}=5.02$ TeV.  Red points are from ALICE collaboration~\cite{Acharya:2018lmh}, colored bands are results
of AMPT calculations. The corresponding cumulant ratios of initial ellipticity $\varepsilon_2$ are shown
as blue squares.}
\label{figapp1}
\end{center}
\end{figure}

\begin{figure}
\begin{center}
\includegraphics[width=0.4\textwidth]{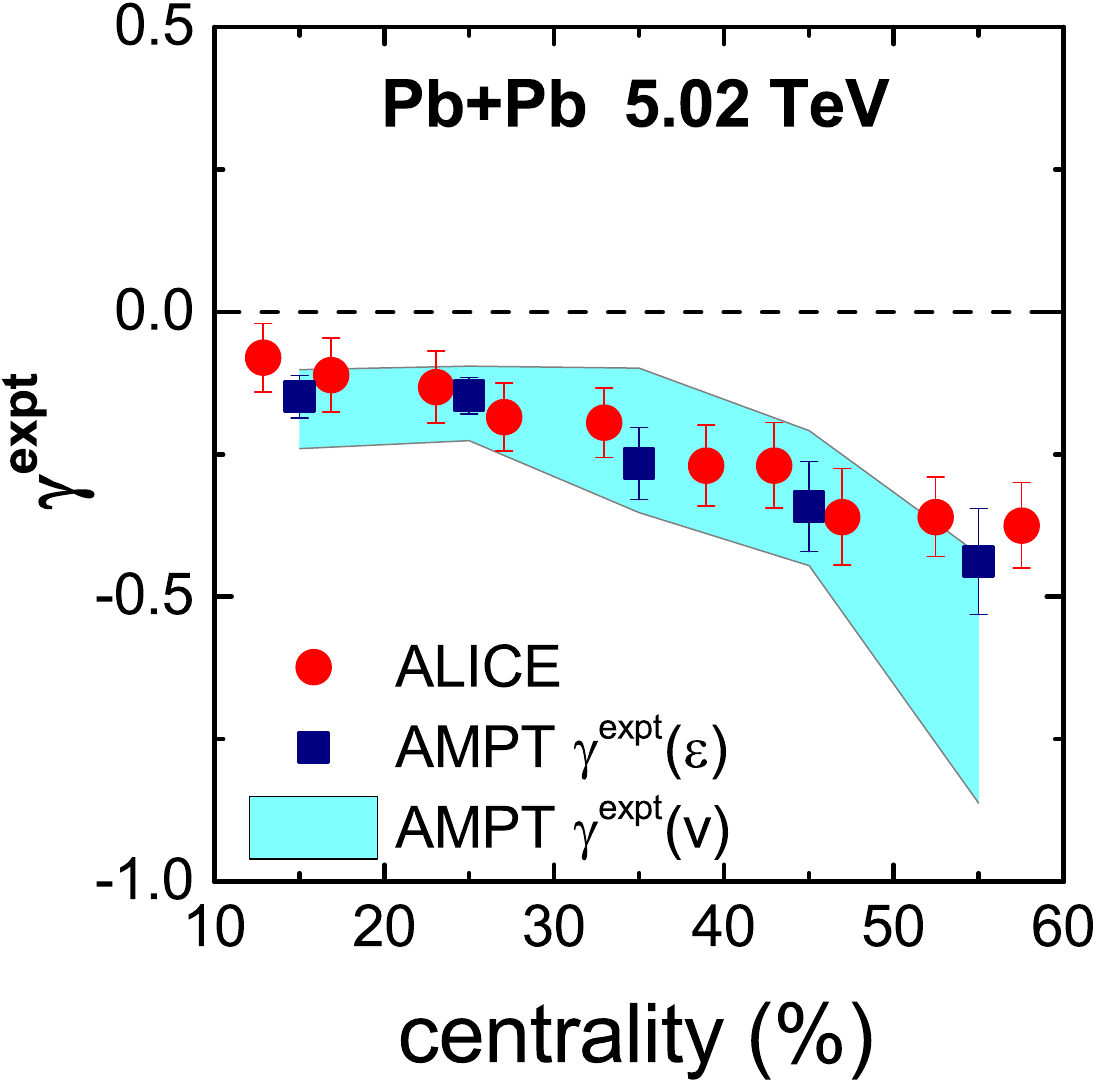}
\caption{(Color online)
Standardized skewness of $v_2$ fluctuations in Pb-Pb collisions at $\sqrt{s_{NN}}=5.02$ TeV as a function of centrality from
AMPT simulations (colored band), and ALICE~\cite{Acharya:2018lmh} data
(red points). The corresponding results of standardized skewness of initial state $\varepsilon_2$
are shown as blue squares. }
\label{figapp2}
\end{center}
\end{figure}

Recent experiments at the LHC has reached $\sqrt{s_{NN}}=5.02$ TeV for Pb-Pb collisions,
where the fluctuations of $v_2$ have been measured.
In this appendix, we present our results of $v_2$ fluctuations in terms of cumulant ratios
in \Fig{figapp1}
and the standardized skewness in
\Fig{figapp2}, from AMPT simulations.
In the AMPT simulations for $\sqrt{s_{NN}}=5.02$ TeV,
we found that the parameter set D (see \Tab{tab1}) can be used to well describe the observed flow signature.
Again, the observed fluctuations of $v_2$ are found compatible with those from initial state
$\varepsilon_2$, implying the dominance of linear response relation between
$v_2$ and $\varepsilon_2$.

\bibliography{paperrefs}
%

\end{document}